\documentclass[preprints,article,accept]{Definitions/mdpi}

\newcommand\arcdeg{\mbox{$^\circ$}}%
\newcommand\arcmin{\mbox{$^\prime$}}%
\newcommand\arcsec{\mbox{$^{\prime\prime}$}}%

\firstpage{1}
\makeatletter
\pubvolume{xx}
\issuenum{1}
\articlenumber{5}
\pubyear{2019}
\copyrightyear{2019}
\history{Received: date; Accepted: date; Published: date}

\Title{Revisit the fraction of radio-loud narrow line Seyfert 1 galaxies with LoTSS DR1}

\Author{Xu-Liang Fan $^{1}$\orcidA{0000-0003-0988-9910}}

\AuthorNames{Xu-Liang Fan}

\address{%
$^{1}$ \quad School of Mathematics, Physics and Statistics, Shanghai University of Engineering Science, Shanghai 201620, People's Republic of China; fanxl@sues.edu.cn}

\abstract{Radio-Loud narrow-line Seyfert 1 galaxies (NLS1s), especially the extremely radio-loud ones, are widely accepted as the jetted versions of NLS1s. We explore the radio-loud fraction for NLS1s with recently released LoTSS DR1 at 150 MHz. The radio detection rate is about 28\% for LoTSS DR1. The radio detected NLS1s have lower redshift than the non-detected ones. Moreover, the 150 MHz radio luminosity of NLS1s detected by LoTSS are about two orders of magnitude weaker than that of the previous samples. By defining the radio loudness with the ratio between 150 MHz radio flux and SDSS \textit{r} band flux, the radio-loud fraction is about 1\% with the critical radio loudness equalling to 100. Radio loudness shows no dependence on central black hole mass, while weak correlations are found between radio loudness and disk luminosity, as well as Eddington ratio. }


\begin{document}
\let\jnl@style=\rmfamily
\def\ref@jnl#1{{\jnl@style#1}}%
\newcommand\aj{\ref@jnl{AJ}}
\newcommand\araa{\ref@jnl{ARA\&A}}
\newcommand\apj{\ref@jnl{ApJ}}
\newcommand\apjl{\ref@jnl{ApJL}}     
\newcommand\apjs{\ref@jnl{ApJS.}}
\newcommand\ao{\ref@jnl{ApOpt}}
\newcommand\apss{\ref@jnl{Ap\&SS}}
\newcommand\aap{\ref@jnl{A\&A}}
\newcommand\aapr{\ref@jnl{A\&A~Rv}}
\newcommand\aaps{\ref@jnl{A\&AS}}
\newcommand\azh{\ref@jnl{AZh}}
\newcommand\baas{\ref@jnl{BAAS}}
\newcommand\icarus{\ref@jnl{Icarus}}
\newcommand\jrasc{\ref@jnl{JRASC}}
\newcommand\memras{\ref@jnl{MmRAS}}
\newcommand\mnras{\ref@jnl{MNRAS}}
\newcommand\pra{\ref@jnl{PhRvA}}
\newcommand\prb{\ref@jnl{PhRvB}}
\newcommand\prc{\ref@jnl{PhRvC}}
\newcommand\prd{\ref@jnl{PhRvD}}
\newcommand\pre{\ref@jnl{PhRvE}}
\newcommand\prl{\ref@jnl{PhRvL}}
\newcommand\pasp{\ref@jnl{PASP}}
\newcommand\pasj{\ref@jnl{PASJ}}
\newcommand\qjras{\ref@jnl{QJRAS}}
\newcommand\skytel{\ref@jnl{S\&T}}
\newcommand\solphys{\ref@jnl{SoPh}}
\newcommand\sovast{\ref@jnl{Soviet~Ast.}}
\newcommand\ssr{\ref@jnl{SSRv}}
\newcommand\zap{\ref@jnl{ZA}}
\newcommand\nat{\ref@jnl{Nature}}
\newcommand\iaucirc{\ref@jnl{IAUC}}
\newcommand\aplett{\ref@jnl{Astrophys.~Lett.}}
\newcommand\apspr{\ref@jnl{Astrophys.~Space~Phys.~Res.}}
\newcommand\bain{\ref@jnl{BAN}}
\newcommand\fcp{\ref@jnl{FCPh}}
\newcommand\gca{\ref@jnl{GeoCoA}}
\newcommand\grl{\ref@jnl{Geophys.~Res.~Lett.}}
\newcommand\jcp{\ref@jnl{JChPh}}
\newcommand\jgr{\ref@jnl{J.~Geophys.~Res.}}
\newcommand\jqsrt{\ref@jnl{JQSRT}}
\newcommand\memsai{\ref@jnl{MmSAI}}
\newcommand\nphysa{\ref@jnl{NuPhA}}
\newcommand\physrep{\ref@jnl{PhR}}
\newcommand\physscr{\ref@jnl{PhyS}}
\newcommand\planss{\ref@jnl{Planet.~Space~Sci.}}
\newcommand\procspie{\ref@jnl{Proc.~SPIE}}
\newcommand\actaa{\ref@jnl{AcA}}
\newcommand\caa{\ref@jnl{ChA\&A}}
\newcommand\cjaa{\ref@jnl{ChJA\&A}}
\newcommand\jcap{\ref@jnl{JCAP}}
\newcommand\na{\ref@jnl{NewA}}
\newcommand\nar{\ref@jnl{NewAR}}
\newcommand\pasa{\ref@jnl{PASA}}
\newcommand\rmxaa{\ref@jnl{RMxAA}}
\newcommand\maps{\ref@jnl{M\&PS}}
\newcommand\aas{\ref@jnl{AAS Meeting Abstracts}}
\newcommand\dps{\ref@jnl{AAS/DPS Meeting Abstracts}}
\section{Introduction}
The formation of jets in active galactic nuclei (AGNs) and their connection with accretion disks are important open questions~\citep{2019ARA&A..57..467B}. Narrow-line Seyfert 1 galaxies/quasars (NLS1s) catch many attentions since their discoveries at $\gamma$-ray band during \textit{Fermi} era~\citep{2009ApJ...707L.142A}, which makes them an important population of jetted AGNs~\citep{2008ApJ...685..801Y,2015A&A...575A..13F}. NLS1s are characterized by their narrow profiles of the broad components of Balmer lines, and high strength of Fe [{\sc ii}] multiple complex~\cite{1985ApJ...297..166O}. Their central engine are believed to be powered by lighter black hole ($<M_{BH}> = 10^{6.5} M_{\odot}$) and higher accretion rate ($<L_{bol}/L_{Edd}> = 0.79$)~\cite{2002ApJ...565...78B, 2012AJ....143...83X}. The radio structures of NLS1s are also found to be compact~\cite{2015ApJS..221....3G}. These features indicate evolved central engines and jet activities of NLS1s, which makes jetted NLS1s useful to understand jet formation and AGN unification~\citep{2017FrASS...4....6F}.

Radio loudness is one of the most common features to find candidates of jetted AGNs~\citep{2017A&ARv..25....2P}. The initial definition of radio loudness is based on the flux ratio between 5 GHz radio band and optical B band (4400 \AA)~\cite{1989AJ.....98.1195K}. Several alternative definitions with different radio or optical bands are also applied in the literature~\citep{2002AJ....124.2364I, 2006AJ....132..531K, 2007ApJ...656..680J, 2017ApJS..229...39R, 2019A&A...622A..11G}.  The scenario that AGNs contain two distinct populations, radio-loud and radio-quiet, is based on the bimodal distribution of radio loudness. Radio emission of radio-loud AGNs is believed to be produced by jets. Meanwhile, many researches suggested that the distribution of radio loudness for AGNs is not bimodal~\citep{2019A&A...622A..11G}. The radio emission originated from star formation activity can also make AGNs appearing as radio-loud~\citep{2015MNRAS.451.1795C, 2017NatAs...1E.194P, 2019A&A...630A.110G}. The combined contribution from jet and star formation makes a continuous distribution of radio loudness. However, for the extreme radio-loud AGNs, i.e., AGNs with radio loudness larger than 100, their radio emission could still be dominated by jet activities~\citep{2019A&A...622A..11G}.

The radio-loud fraction of NLS1s is found to be smaller than that of other jetted AGNs.~\citeauthor{2006ApJS..166..128Z} (\citeyear{2006ApJS..166..128Z}) built the largest NLS1 sample at the time with SDSS DR3. They found that the fraction of NLS1s decreases as the radio loudness increases.~\citeauthor{2006AJ....132..531K} (\citeyear{2006AJ....132..531K}) also explored a sample of optically selected NLS1s and found the radio-loud fraction is about 7\%, lower than the typical 10\% - 20\% for quasars~\citep{2007ApJ...656..680J,2016ApJ...831..168K}. Recently, ~\citeauthor{2018MNRAS.480.1796S} (\citeyear{2018MNRAS.480.1796S}) explored the radio associations of a sample of optically selected NLS1s with several radio surveys. The radio detection rate of NLS1s were found to be low, from 0.7 percent to 4.6 percent. However, majority of radio-detected NLS1s could be defined as radio-loud\cite{2018MNRAS.480.1796S}.

Moreover, the extreme radio-loud NLS1s, which have similar features with blazars, are widely explored in the literature~\cite{2015A&A...575A..13F}. According to the unification model, there expected to find 2$\Gamma^2$ ($\Gamma$ is the bulk Lorentz factor of jet, which is about 10 in $\gamma$-ray NLS1s\cite{2019ApJ...872..169P}) misaligned radio-loud NLS1s~\cite{2016A&A...591A..98B}. However, the steep-spectrum radio-loud NLS1s are found to be rare than the prediction of unification model~\cite{2015A&A...578A..28B}. One possibility to explain this is that there are fractional radio-loud NLS1s missed in the previous radio surveys. Therefore, it is important to built samples of radio-loud NLS1s with more sensitive radio surveys.

Radio radiation of AGN jets is usually dominating at low radio frequency, results in a steep radio spectra. Thus the low-frequency radio survey is expected to be more efficient to detect jetted AGNs. However, the sensitivities of low-frequency radio surveys are much higher than those at higher radio frequencies, such as the FIRST survey. Recent years, several low-frequency radio surveys has been performed and released their source catalogues~\citep{2017A&A...598A..78I, 2017MNRAS.464.1146H,2019A&A...622A...1S}. Their sensitivity and sample sizes are comparable, even better than the radio surveys at GHz band~\citep{2019A&A...622A...1S}. Thus it is feasible to examine the radio-loud fraction of NLS1s at low radio band.

In this paper, we revisit the radio detection rate and radio-loud fraction of NLS1s with the first data release of LOFAR Two-metre Sky Survey (LoTSS DR1)~\citep{2019A&A...622A...1S}. In Section 2, the method for sample selection in our work is described. Section 3 explore the properties of radio detected NLS1s, and the nature of their radio loudness. In section 4, we compare our results with previous researches, and discuss the origin of radio emission for the LoTSS detected NLS1s. The main results are summarized in section 5.

\section{Sample}
The LOw Frequency ARray (LOFAR;~\citeauthor{2013A&A...556A...2V} ~\citeyear{2013A&A...556A...2V}) constructed in the north of the Netherlands and across europeis is a radio interferometer covered the low-frequency radio band from 10-240 MHz. LoTSS is a 120 - 168 MHz radio survey with the High
Band Antenna (HBA) system of LOFAR. Making use of direction-dependent calibration, the first full quality public data release of LoTSS (LoTSS DR1) reaches the median sensitivity 71 $\mu$Jy beam$^{-1}$, the resolution 6\arcsec and the positional accuracy within 0.2\arcsec. 424 square degrees in the region of HETDEX Spring Field (right ascension 10h45m00s to 15h30m00s and declination 45\arcdeg~00\arcmin~00\arcsec~to
57\arcdeg~00\arcmin~00\arcsec~) are mapped in LoTSS DR1, which contains 325 694 sources with $\sigma > 5$~\citep{2019A&A...622A...1S}. \citeauthor{2019A&A...622A...2W} (\citeyear{2019A&A...622A...2W}) associated LoTSS DR1 with several  optical and IR catalogues, and removed various artefacts due to the limitations of Python Blob Detector and Source Finder (PyBDSF). Then they built the LoTSS-DR1 value-added catalogue (hereafter LoTSS DR1-ID), which contains 318 520 source in total.

\citeauthor{2017ApJS..229...39R} (\citeyear{2017ApJS..229...39R}) modelled the spectra of the quasars in SDSS DR12, and derived a catalog with 11101 NLS1 candidates, in which 968 objects are within the HETDEX Spring Field. As suggested by~\citeauthor{2020CoSka..50..270B} (\citeyear{2020CoSka..50..270B}), there are some spurious non-NLS1 sources in the sample of \citeauthor{2017ApJS..229...39R} (\citeyear{2017ApJS..229...39R}), especially for the sources with low signal-to-noise ratio (S/N). Thus we remove the NLS1 candidates with median S/N less than 5 per pixel in our analysis. This leaves 725 NLS1s in the field of LoTSS DR1. In \citeauthor{2017ApJS..229...39R} (\citeyear{2017ApJS..229...39R}), the authors listed the properties of the emission lines and continuum. Based on the $\lambda 5100$ luminosity and the FWHM of H$\beta$, central black hole mass and disk luminosity can be estimated with several empirical relations~\citep{2011ApJS..194...45S}.

For comparison, we build a sample of broad-line Seyfert 1 galaxies (BLS1s) from the 13th version of the Catalogue of Quasars and Active Nuclei compiled by \citeauthor{2010A&A...518A..10V} (\citeyear{2010A&A...518A..10V}). Among the 13975 sources labelled as "S1", there are 500 objects in the field of LoTSS DR1.

The optical photometric magnitudes of NLS1s and BLS1s are derived from the website of SDSS CrossID for DR12~\footnote{http://skyserver.sdss.org/dr12/en/tools/crossid/crossid.aspx}. For the 500 BLS1s in LoTSS DR1 region, there are 491 associations in SDSS website. The galactic extinctions of all sources are derived from IRSA Dust Extinction Service~\footnote{https://irsa.ipac.caltech.edu/applications/DUST/index.html}, which are based on the extinction law of \citeauthor{2011ApJ...737..103S} (\citeyear{2011ApJ...737..103S}).

The optically selected sample of NLS1s and BLS1s are cross-matched with LoTSS DR1-ID with the radius 5\arcsec. 204 and 181 radio counterparts of NLS1s and BLS1s are found, respectively. The corresponding radio detection rate is about 28.1\% and 36.2\%, respectively. 

\section{Results}
\subsection{The Properties of Radio Detected NLS1s}
\begin{figure}[H]
\centering
\includegraphics[width=10 cm]{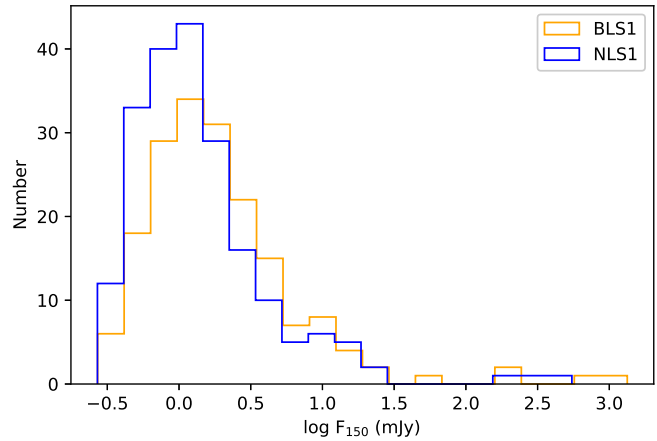}
\caption{The distribution of 150 MHz radio flux for LoTSS detected NLS1s and BLS1s. \label{f151}}
\end{figure}

Figure~\ref{f151} shows the distribution of 150 MHz radio flux for NLS1s and BLS1s. NLS1s and BLS1s show similar ranges of 150 MHz radio flux, while the flux of the brightest BLS1 (1334.17 mJy) is higher than that of the brightest NLS1 (548.39 mJy). The Kolmogorov-Smirnov (K-S) test is applied to evaluate whether the flux distribution of the two samples are drawn from the same distribution~\cite{1992nrfa.book.....P}. We consider that the null hypothesis that two populations are drawn from the same distribution, can not be rejected when the probability is greater than 0.05. The K-S test denies that the flux distributions of NLS1s and BLS1s are drawn from the same sample, with the probability equalling to 0.002. BLS1s show higher mean 150 MHz flux (16.56 mJy) than NLS1s (7.14 mJy).

For NLS1s, we estimate their black hole mass and disk luminosity with the luminosity of $\lambda$5100 and FWHM of H$\beta$. The empirical relations of~\citeauthor{2011ApJS..194...45S} (\citeyear{2011ApJS..194...45S}) are applied with $\log M_{BH} = 0.91 + 0.5\log(L_{5100}/10^{44}) + 2\log(FWHM)$ and $L_{bol} = 9.26 L_{5100}$. Typical uncertainties 0.4 dex and 0.3 dex for black hole mass and disk luminosity of these relations are considered, respectively. The Eddington ratio $L_{bol}/L_{Edd}$ is then calculated, where $L_{Edd} = 1.3\times10^{38} M_{BH}/M_{\odot}$ erg s$^{-1}$ is the Eddington luminosity.

\begin{figure}[H]
\centering
\includegraphics[width=7 cm]{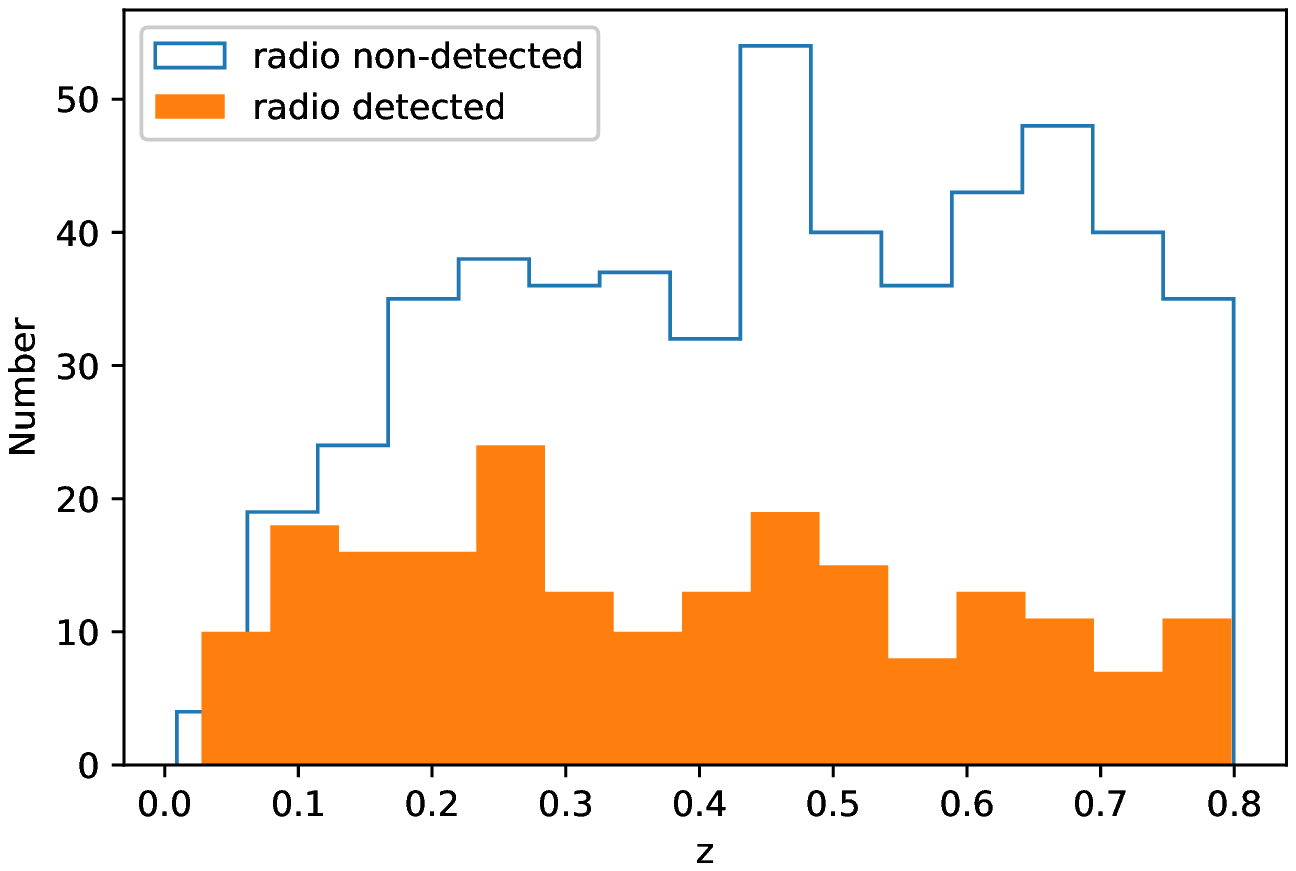}
\includegraphics[width=7 cm]{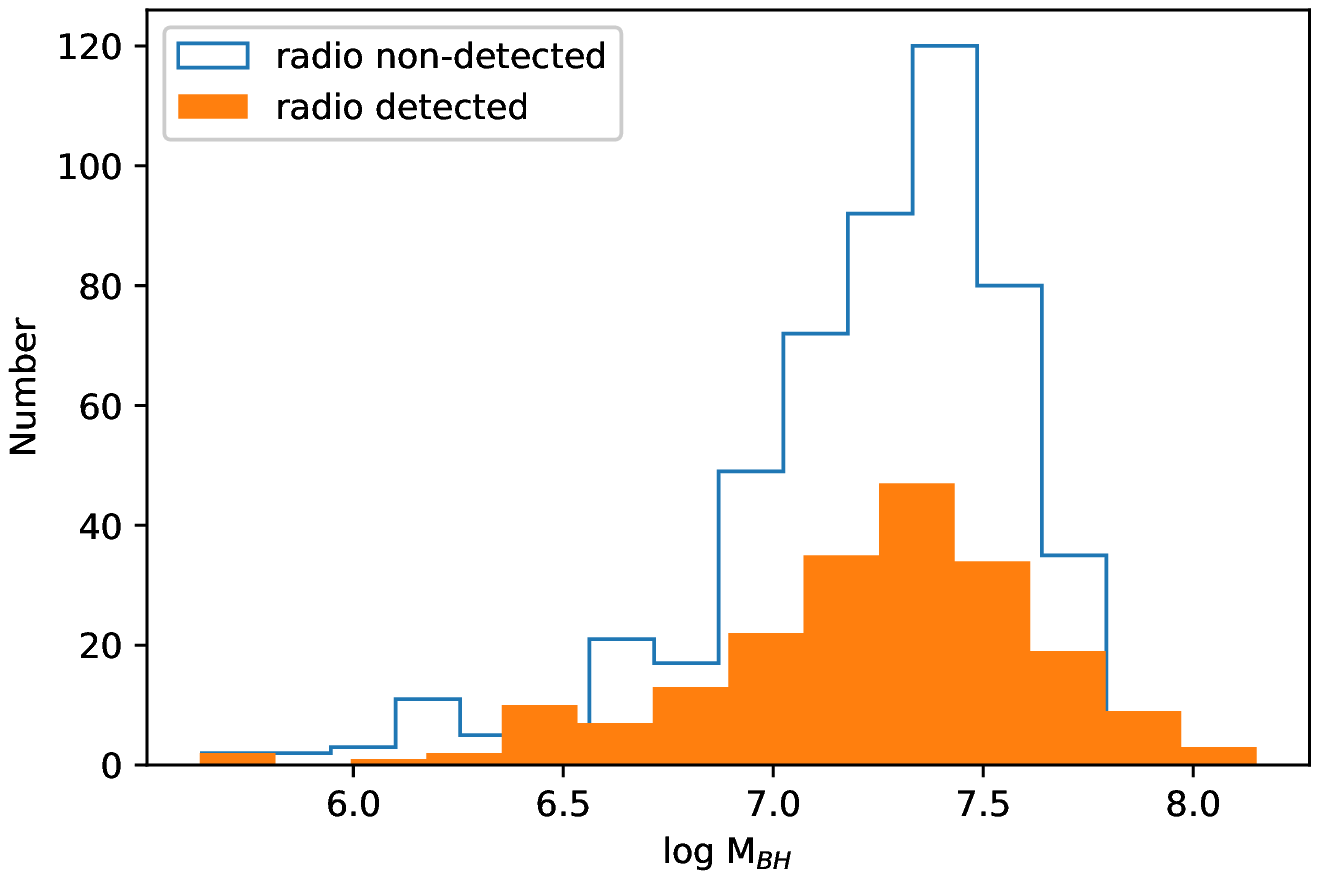}
\includegraphics[width=7 cm]{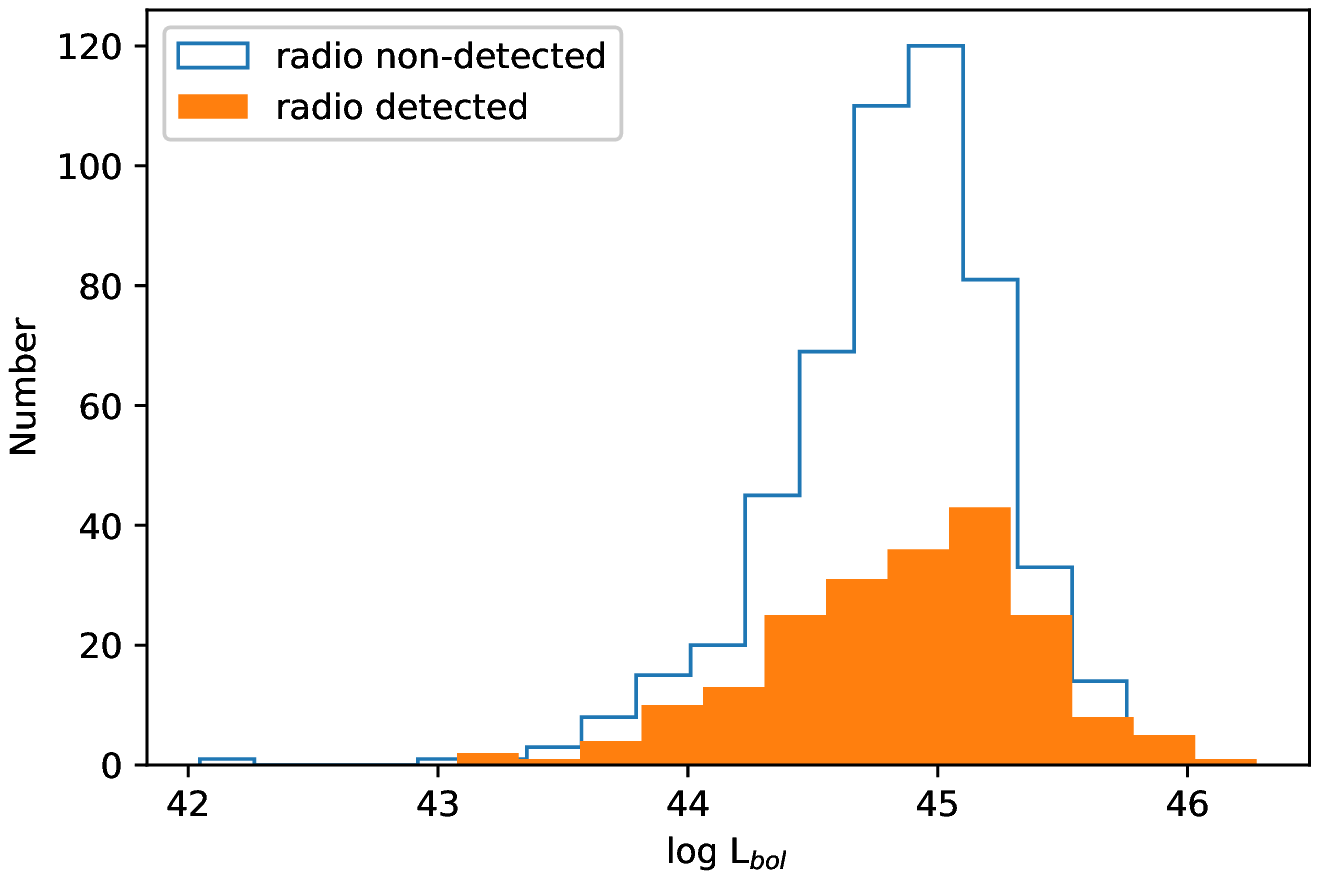}
\includegraphics[width=7 cm]{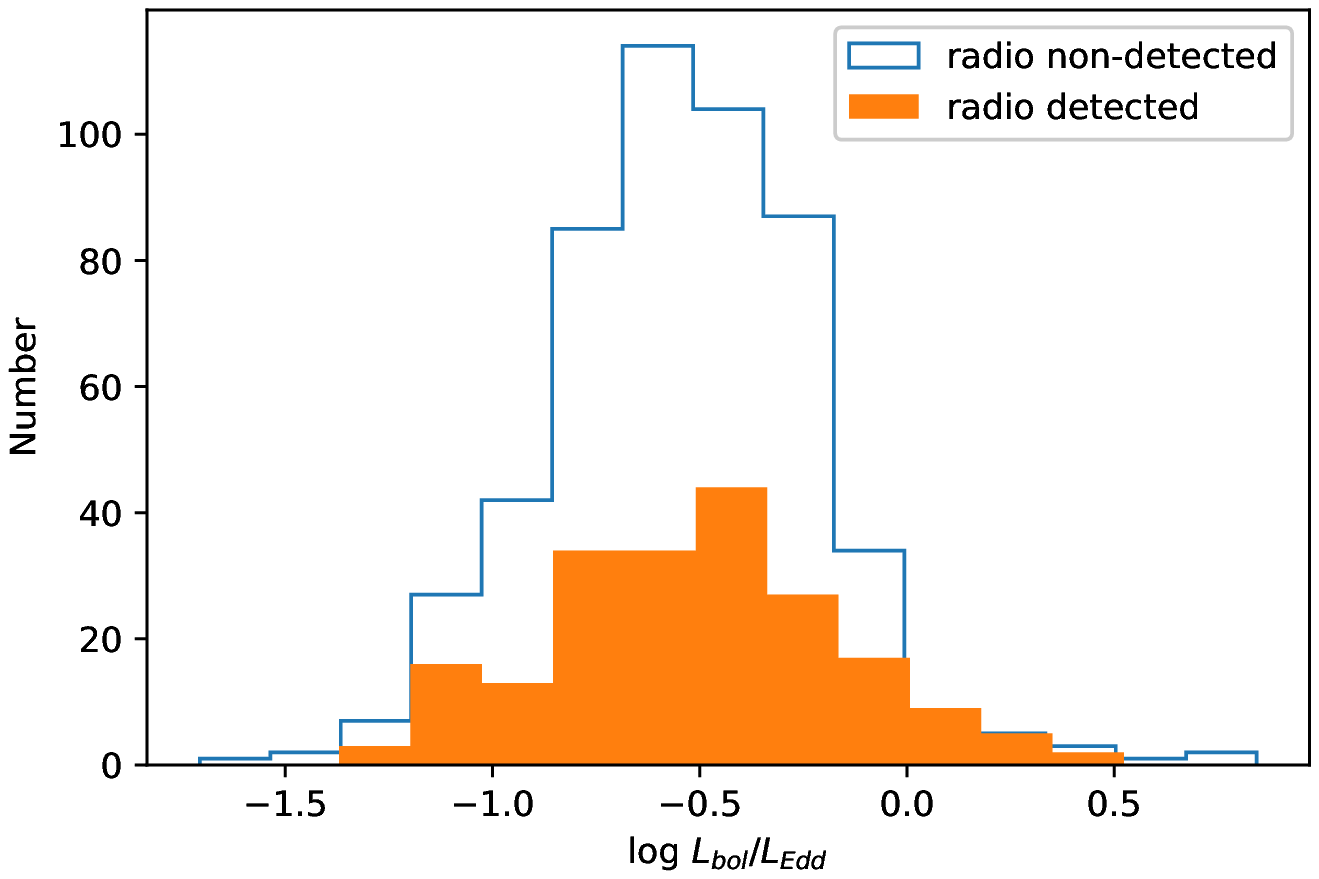}
\caption{Comparisons of the properties between radio detected and non-detected NLS1s. Top left: redshift. Top right: central black hole mass. Bottom left: disk luminosity. Bottom right: Eddington ratio.\label{undetected}}
\end{figure}

\citeauthor{2018MNRAS.480.1796S} (\citeyear{2018MNRAS.480.1796S}) noted that radio detected NLS1s have relatively smaller redshift. In order to clarify this, we compare the redshift distribution between LoTSS detected and non-detected NLS1s (Top left panel of Figure~\ref{undetected}). The K-S test shows that they are drawn from distinct sample with probability of $2\times10^{-4}$. The LoTSS detected sample also show relatively lower redshift with respect to the non-detected one, with the mean value 0.38 compared to 0.46.

We then explore whether there are differences of intrinsic properties between radio detected and non-detected NLS1s (Figure~\ref{undetected}). The distributions of black hole mass, disk luminosity, and Eddington ratio show no obvious difference between radio detected and non-detected objects. The K-S tests confirm that they are drawn from the same distribution with the probability equalling to 0.45, 0.07 and 0.17 for black hole mass, disk luminosity, and Eddington ratio, respectively.

\begin{figure}[H]
\centering
\includegraphics[width=7.5 cm]{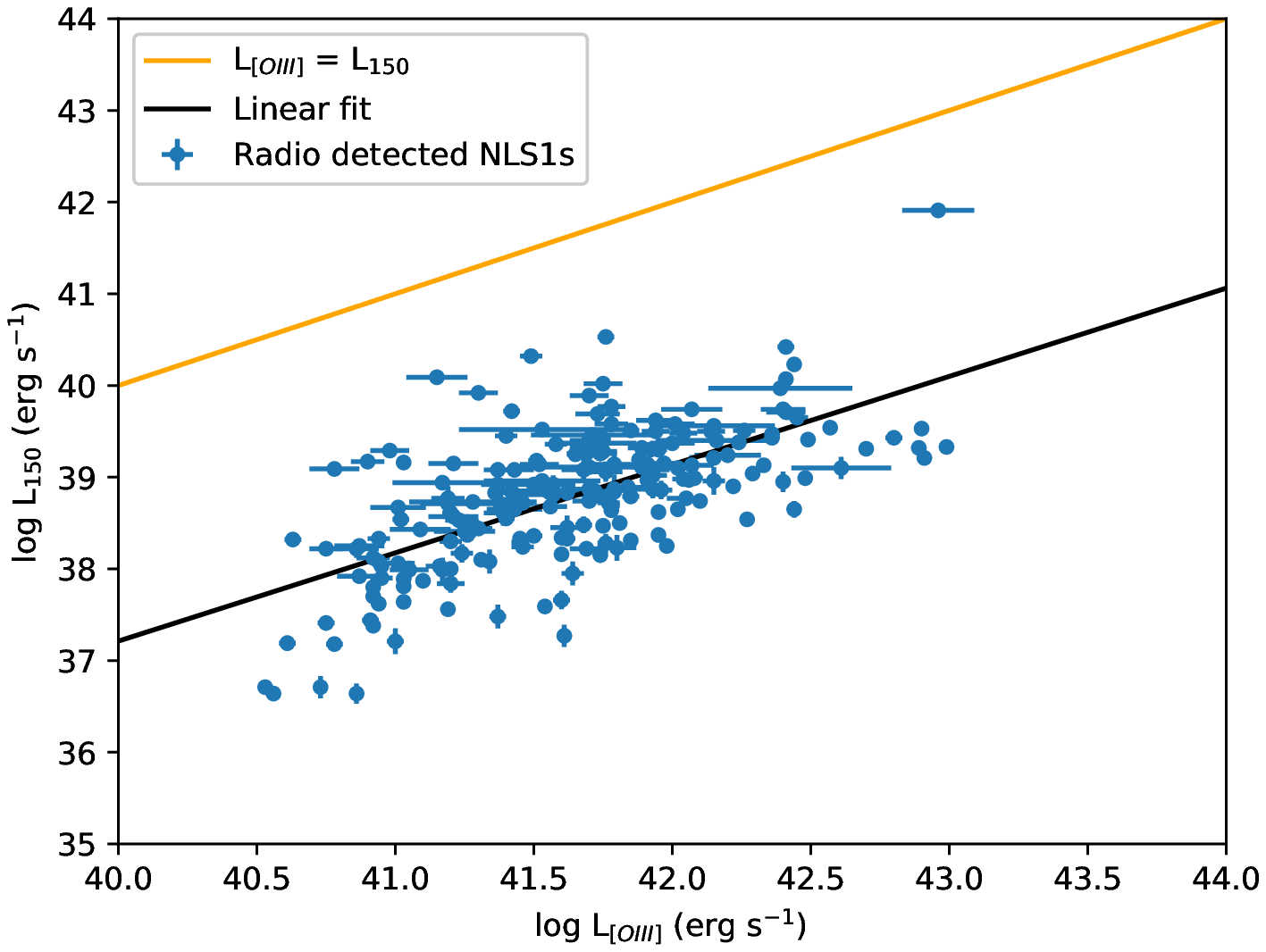}
\includegraphics[width=7.5 cm]{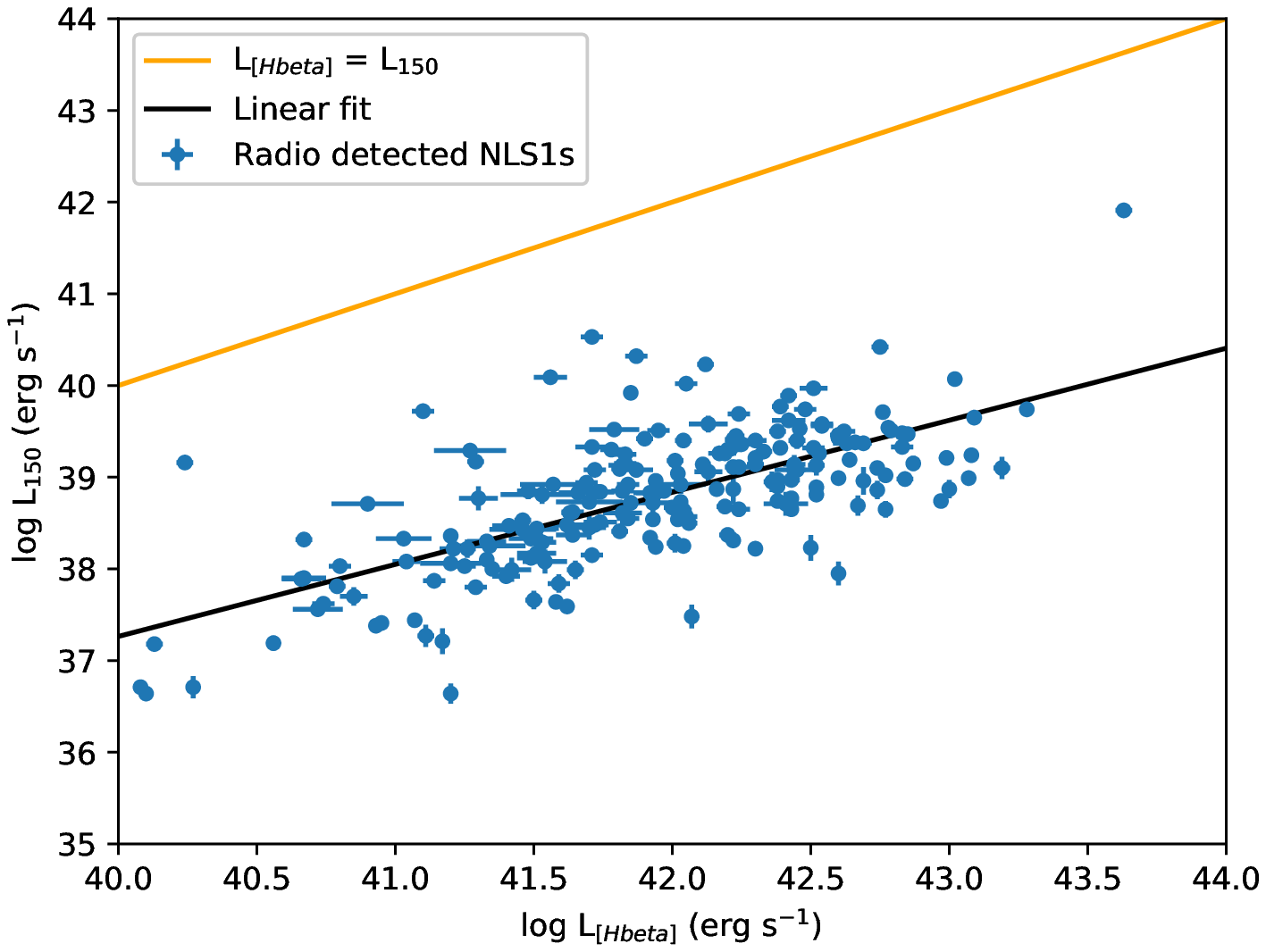}
\caption{Left panel: The connection between [O {\sc iii}] luminosity and 150 MHz radio luminosity.Right panel: The connection between the luminosity of broad H$\beta$ and 150 MHz radio luminosity. The black solid lines show the best fits, while the orange solid lines represent the line where the two parameters are equal with each other. \label{l150}}
\end{figure}

The radio emission of jet are found to be connected with the luminosity of emission lines, such as [O {\sc iii}] and broad H$\alpha$ (e.g., \cite{1999MNRAS.309.1017W, 2010A&A...509A...6B}). These connections are believed to be due to their common energy source from central engine. Thus they are considered as evidences of jet-disk connection. For the radio detected NLS1, we explore the connection between 150 MHz radio luminosity and [O {\sc iii}] luminosity (left panel of Figure~\ref{l150}). In order to exclude the common influences of redshift on luminosity, the partial Kendall's $\tau$ correlation test is performed to examine the luminosity correlation between radio band and emission lines~\citep{1996MNRAS.278..919A}. The null hypothesis of zero partial correlation can be rejected at a significance 0.95 if the ratio between the correlation coefficient and its statistical variance ($\tau/\sigma$) larger than 1.96. The correlation between 150 MHz radio luminosity and [O {\sc iii}] luminosity is confirmed with $\tau = 0.27$ and $\sigma = 0.04$. Then a Bayesian approach of linear regression~\cite{2007ApJ...657..116K} is applied to explore the linear relation between these two parameters. The result gives:
\begin{equation}
\label{oiii}
\log L_{150} = (0.96\pm0.08)\log L_{[O III]} - (1.28\pm3.29)
\end{equation}

As the emission of [O {\sc iii}] can be produced in the star formation region, we also compare the 150 MHz radio luminosity with the luminosity of broad H$\beta$ (right panel of Figure~\ref{l150}). The partial Kendall's $\tau$ correlation test shows that the 150 MHz radio luminosity is correlative with the luminosity of H$\beta$, with $\tau = 0.22$ and $\sigma = 0.04$. The linear fit shows a relatively flatter slope with
\begin{equation}
\label{hb}
\log L_{150} = (0.79\pm0.06)\log L_{H\beta} + (5.83\pm2.51)
\end{equation}

\subsection{Radio Loudness}
\begin{figure}[H]
\centering
\includegraphics[width=10 cm]{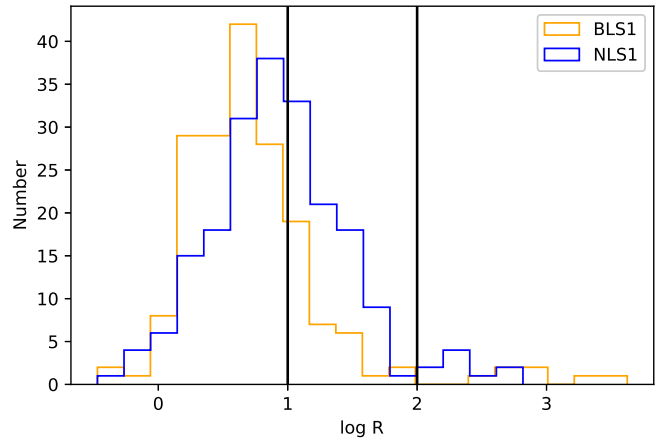}
\caption{The distribution of radio loudness for LoTSS detected NLS1s and BLS1s. The vertical lines show the radio loudness equalling to 10 and 100, respectively. \label{loudness}}
\end{figure}

Radio loudness can be used as a tracer of jet activity based on the assumption that radio emission is dominated by the non-thermal radiation from jet, while the optical emission can trace the activity of accretion process of central black hole~\citep{2013ApJ...764L..24S}. The radiation of accretion disk is dominating at UV band. That is why emission at short wavelength are favored to estimate radio loudness. Meanwhile, at blue band the optical emission can be diluted by the young star formation, which is important in star-forming galaxy and also contributes to the radio emission~\citep{2019NatAs...3..387P}. Blue band also suffers more dust extinction. Thus SDSS \textit{g} band and \textit{i} band are both used to define radio loudness\citep{2017ApJS..229...39R, 2019A&A...622A..11G}. Here, we use the flux ratio between 150 MHz radio band and SDSS \textit{r} band, which is between \textit{g} band and \textit{i} band. The distributions of radio loudness for NLS1s and BLS1s are presented in Figure~\ref{loudness}.

The distribution of radio loudness for both NLS1s and BLS1s show no clear bimodal, although a tail at high end is plausible. The classical threshold between radio-loud and radio-quiet is 10 with the definition of radio loudness as the ratio between 5 GHz and 4400 \AA~flux~\cite{1989AJ.....98.1195K}. It can be converted into our definition of radio loudness with the spectral index $\alpha$~\footnote{$F_{\nu} \propto \nu^{-\alpha}$} =  0.7 and 0.5 for radio and optical band, respectively~\cite{2006AJ....132..531K}. Taken 150 GHz and the central wavelength of \textit{r} 6166 \AA~\footnote{https://www.sdss.org/instruments/camera/\#Filters} into calculation, the threshold becomes $\sim100$. In this work, we calculate the radio-loud fraction with both thresholds of 10 and 100. Among the 204 NLS1s, there are 86 ones with radio loudness larger than 10, and only 9 ones with radio loudness larger than 100. The radio-loud fractions are 11.9\% and 1.2\%, respectively. For 181 BLS1s, 38 objects have radio loudness larger than 10, and 7 ones larger than 100. Radio-loud fraction of BLS1s is 7.6\% for critical radio loudness 10, and 1.4\% for critical value 100.

\begin{figure}[H]
\centering
\includegraphics[width=7.5 cm]{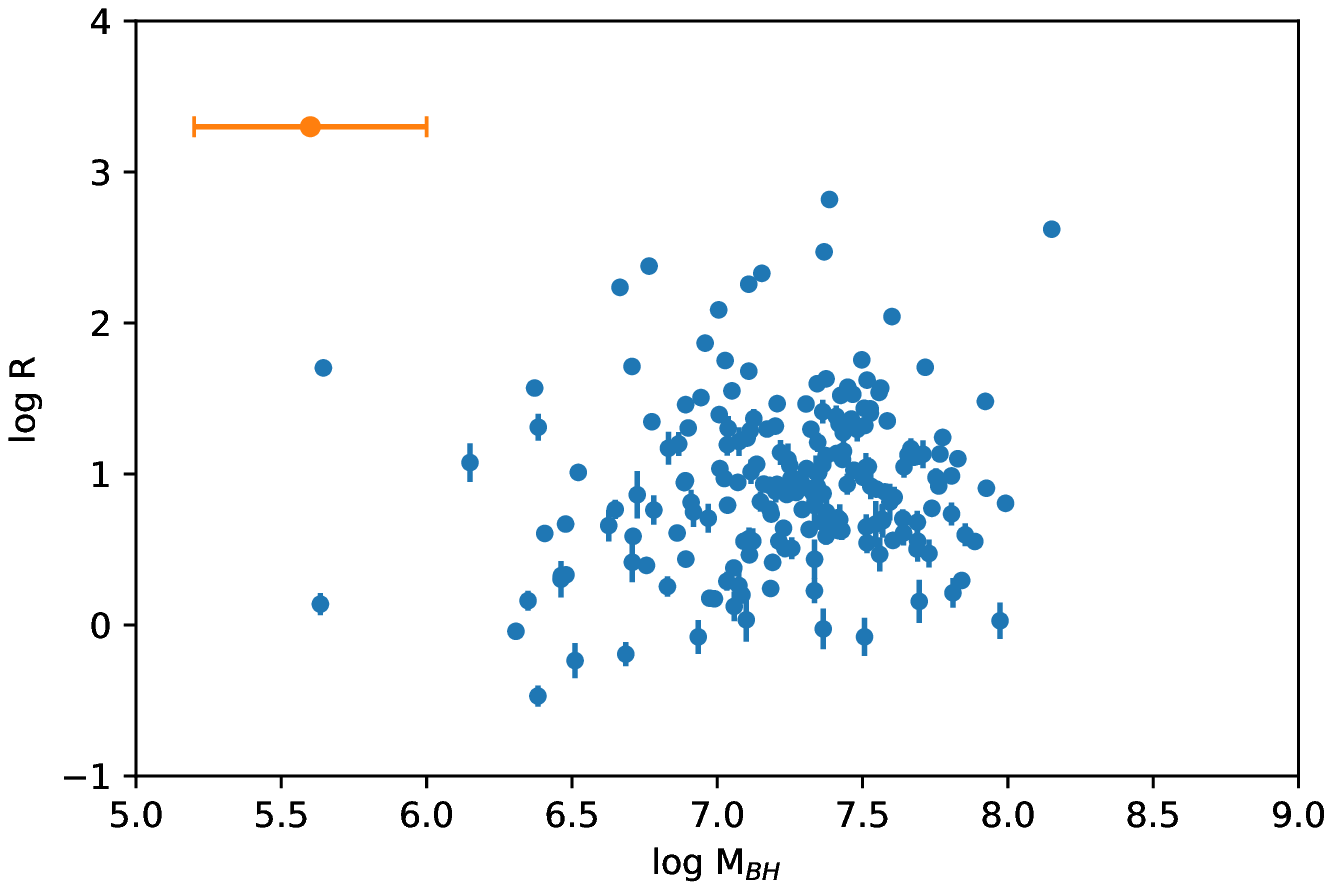}
\includegraphics[width=7.5 cm]{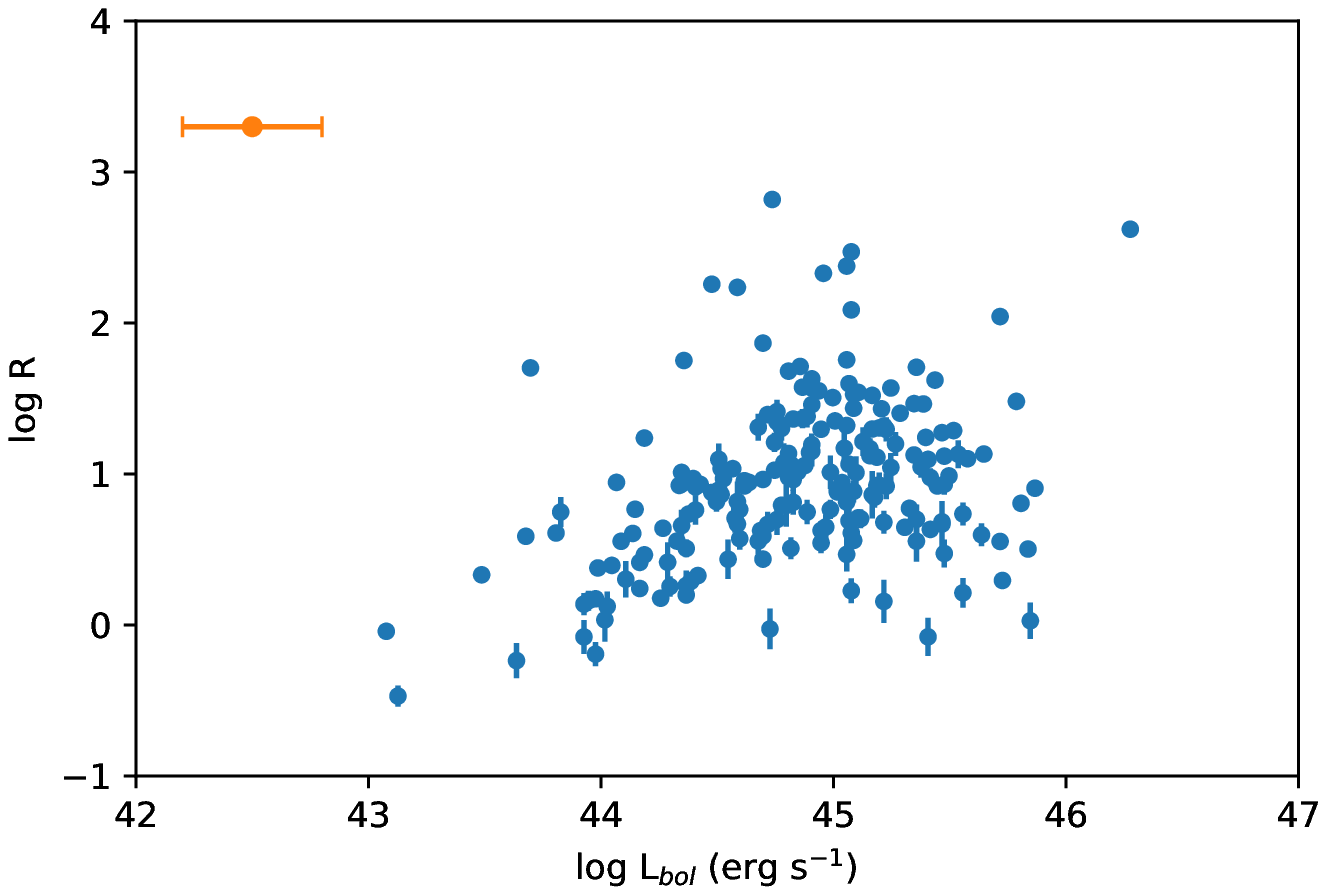}
\includegraphics[width=7.5 cm]{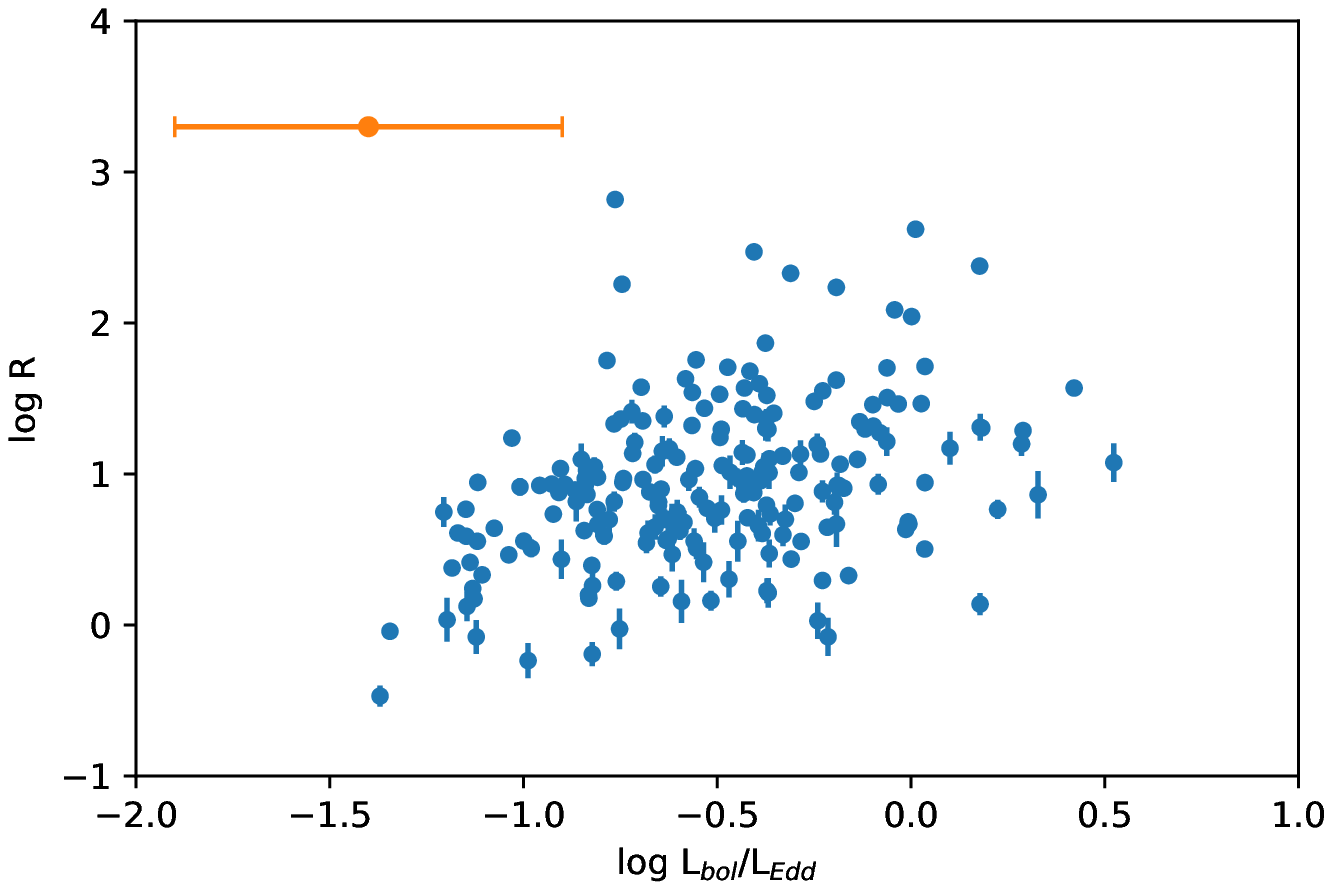}
\caption{The dependence of radio loudness on the black hole mass (top left panel), disk luminosity (top right panel), and Eddington ratio (bottom panel). On the top left corner of each panel, the typical uncertainties of black hole mass, disk luminosity and Eddington ratio are labelled, respectively. \label{rl_cent}}
\end{figure}

The radio loudness is found to show possible correlations with black hole mass and Eddington ratio~\cite{2000ApJ...543L.111L, 2019A&A...622A..11G}. We examine the correlation between radio loudness of NLS1s and black hole mass, as well as other properties of central activities (Figure~\ref{rl_cent}). The Spearman rank-order correlation test are employed to examine the correlation between them. We consider a possible correlation when the chance probability of a correlation is less than 0.05. No correlation is found between radio loudness and central black hole mass with the correlation coefficient $\rho = 0.11$ and the chance probability $P = 0.13$. Weak correlation is found between radio loudness and disk luminosity with $\rho = 0.33$ and $P = 1.5\times10^{-6}$. Moreover, radio loudness is correlated with Eddington ratio, with $\rho = 0.38$ and $P = 2.6\times10^{-8}$.

\section{Discussion}
Since the discovery of $\gamma$-ray emission from NLS1s, they are believed to host relativistic jets~\citep{2017FrASS...4....6F}. Radio loudness is a widest used probe to find jetted AGNs. The radio-loud fraction of NLS1s is found to be lower than other types of AGNs, such as broad line Seyfert galaxies, or quasars~\cite{2006AJ....132..531K}. Here, we use 150 MHz radio survey LoTSS DR1-ID to explore the radio properties for a optically selected sample of NLS1s from SDSS DR12. The detection rate is about 28\%, while the radio-loud fraction is about 12\% and 1\% for the threshold between radio-loud and radio-quiet equalling to 10 and 100, respectively. For comparison, the detection rate of LoTSS detected BLS1s is 36\%. The radio-loud fraction of BLS1s is 8\%, smaller than that of NLS1s, when 10 is taken as the critical radio loudness. For the critical radio loudness of 100, the radio-loud fraction is similar between NLS1s and BLS1s.

The detection rate of NLS1s with LoTSS DR1 is much higher than that at GHz band, as well as NLS1s detected by TIFR Giant metrewave radio telescope Sky Survey (TGSS) at 150 MHz. \citeauthor{2017ApJS..229...39R} (\citeyear{2017ApJS..229...39R}) searched the radio counterparts with FIRST 1.4 GHz radio survey for the same optically selected sample of NLS1s. They only find 555 radio associations. The radio detection rate is about 5\%.  \citeauthor{2018MNRAS.480.1796S} (\citeyear{2018MNRAS.480.1796S}) associated NLS1s with TGSS. They got a very low detection rate of 0.7\%. The median flux of NLS1s detected by TGSS is $\sim$ 100 mJy, while most LoTSS detected NLS1s have 150 MHz radio flux fainter than 10 mJy (Figure~\ref{f151}). Similar with the results of \citeauthor{2018MNRAS.480.1796S} (\citeyear{2018MNRAS.480.1796S}), we also find that radio NLS1s can be effectively detected at lower redshift.

The higher detection rate of NLS1s and their comparable radio-loud fraction with BLS1s indicate that the radio properties become very different, when the sensitivities of radio surveys get better, and more and more fainter AGNs are discovered. The median of 150 MHz radio luminosity of TGSS detected NLS1s is $10^{40.97}$ erg s$^{-1}$~\cite{2018MNRAS.480.1796S}. The 150 MHz radio luminosity of LoTSS detected NLS1s ranges from $10^{36.64}$ erg s$^{-1}$ to $10^{41.91}$ erg s$^{-1}$, with the mean value $10^{38.79}$ erg s$^{-1}$. The radio luminosity of LoTSS detected NLS1s is about two orders of magnitude weaker than that of TGSS ones. As the strength of radio emission get weaker, the origin of radio emission for these radio NLS1s gets complicated. Relativistic jets, star formation activity, AGN-driven wind, as well as corona associated with accretion disk can contribute to the radio emission~\citep{2019NatAs...3..387P}. The results of the radio/Far-IR correlation also suggests the radio emission in low luminosity optically selected quasars could be dominated by star formation activities~\cite{2019A&A...622A..11G}.  \citeauthor{2015MNRAS.451.1795C} (\citeyear{2015MNRAS.451.1795C}) investigated the mid-IR properties for a sample of radio-loud NLS1s with flat radio spectra~\cite{2015A&A...575A..13F}. They concluded that star formation activities could give important contributions at both radio and mid-IR bands.~\citeauthor{2018MNRAS.480.1796S} (\citeyear{2018MNRAS.480.1796S}) also explored the star formation contribution of radio detected NLS1s with mid-IR data of WISE. They found that NLS1s with low radio luminosity show similar ratio between IR and radio flux ($q$) with luminous IR galaxies. But they also cautioned that the mid-IR emission is dominated by AGN activities.

In this work, we explore several correlations between jet and accretion activities to clarify the origin of radio emission for LoTSS detected NLS1s. The radio-emission line connections are common found in radio-loud AGNs, which are explained by the common energy source for both optical line and radio emission~\cite{1999MNRAS.309.1017W, 2010A&A...509A...6B}. The slope of the linear relation of $L_{[O III]}$ dependent on $L_{radio}$ is around unity~\cite{2010A&A...509A...6B}.~\citeauthor{2017ApJS..229...39R} (\citeyear{2017ApJS..229...39R}) explored the correlation between 1.4 GHz radio luminosity and [O {\sc iii}] luminosity for FIRST detected NLS1s. They got $L_{1.4} \propto L_{[O III]}^{0.98}$, which is well consistent with our finding in Equation~\ref{oiii}. We also examine the correlation between $L_{150}$ and $L_{H\beta}$. This correlation is also confirmed by the partial Kendall's $\tau$ correlation test. Star formation activity can contribute both radio and narrow line emission, but broad emission line is believed to be powered by the accretion power of AGN. Radio emission from star formation activity is not expected to be correlated with emission of broad H$\beta$. Thus the luminosity correlation between 150 MHz and broad H$\beta$ indicates that radio emission of LoTSS detected NLS1s is still dominated by jet activity.

The dependence of radio loudness on black hole mass was presented to support that jet is more frequent in heavier black hole system~\cite{2000ApJ...543L.111L}. In addition, radio loudness is suggested to be negatively correlative with Eddington ratio, due to the transition of accretion mode~\cite{2002ApJ...564..120H, 2007ApJ...658..815S}. For LoTSS detected NLS1s, no correlation is found between radio loudness and black hole mass, while positive correlation is found between radio loudness and Eddington ratio. The main reason of the different results for Eddington ratio is that NLS1s are accreting with much higher accretion rate ($L_{bol}/L_{Edd} > 0.03$).~\citeauthor{2012A&A...545A..66B} (\citeyear{2012A&A...545A..66B}) dealt with a sample of type 1 AGNs, where most sources had $L_{bol}/L_{Edd} > 0.01$, and found radio loudness is positively correlative with Eddington ratio. The connection between radio loudness and Eddington ratio also indicates that the radio emission of LoTSS detected NLS1s is controlled by the accretion activity. However, these evidences are still indirect. In order to clarify the origin of their radio emission, the multi-wavelength properties of these low radio luminosity NLS1s need to be explored in the future, such as the radio/Far-IR correlation~\citep{2017NatAs...1E.194P} and multi-band SEDs.

\begin{figure}[H]
\centering
\includegraphics[width=7.5 cm]{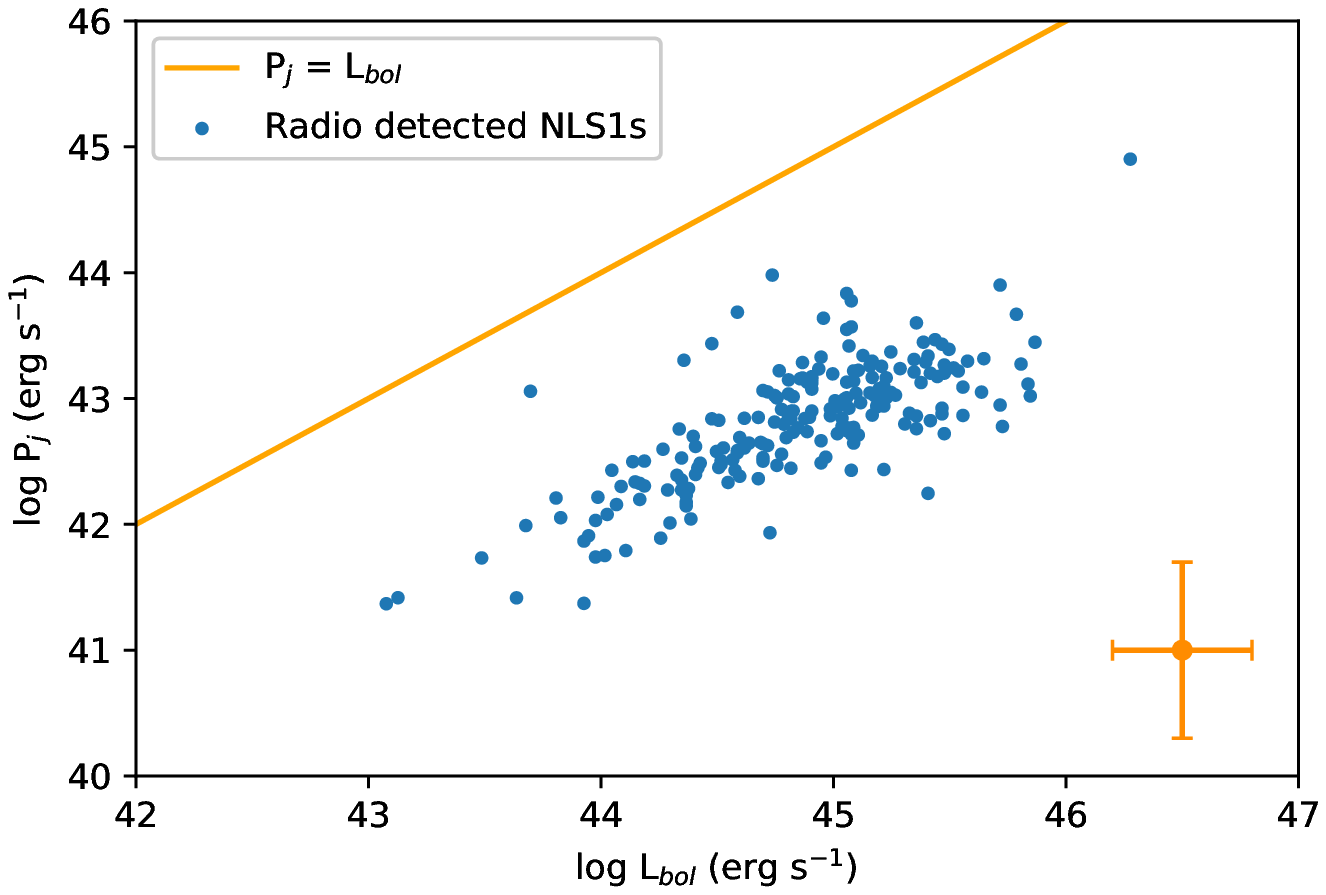}
\includegraphics[width=7.5 cm]{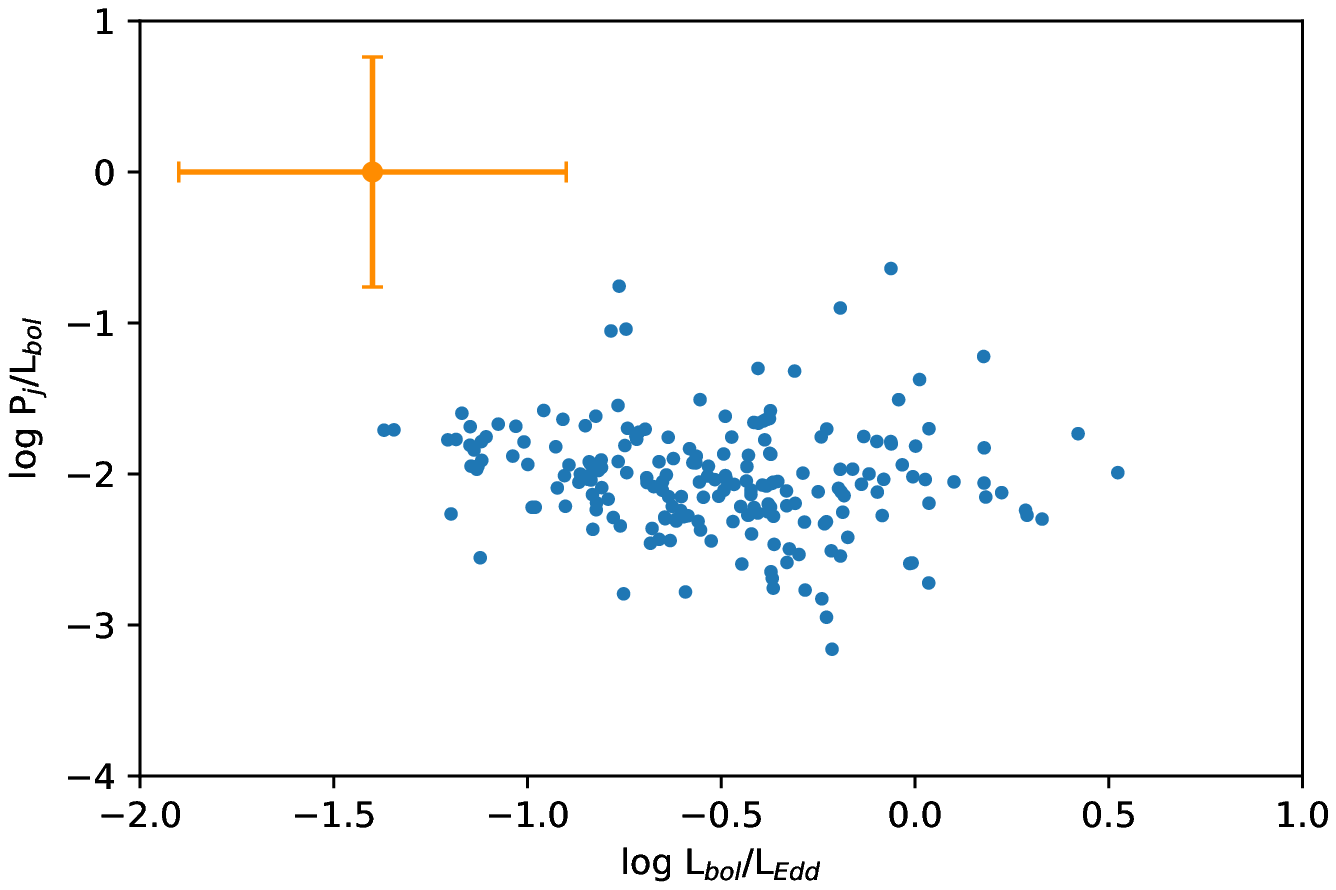}
\caption{Left panel: The connection between disk luminosity and jet power. The solid line represents the line where the two parameters are equal with each other. Right panel: The connection between Eddington ratio and jet production efficiency. The typical uncertainties are labelled on each panel. \label{jet-disk}}
\end{figure}

Although radio emission originated from star formation are suggested for AGNs with low radio luminosity~\cite{2016A&ARv..24...13P, 2019A&A...622A..11G}, there are several indirect evidences support that the low-frequency radio emission of LoTSS detected NLS1s is related to the AGN activities, not the star formation activities. Therefore, following \citeauthor{2019ApJ...879..107F} (\citeyear{2019ApJ...879..107F}), we estimate the jet power for the LoTSS detected NLS1s assuming that radio emission originates from jet. The typical uncertainty of jet power is taken as 0.7 dex~\cite{2010ApJ...720.1066C}. Then we explore the jet-disk connection of LoTSS detected NLS1s. The left panel of Figure~\ref{jet-disk} shows the connection between the disk luminosity and jet power. The partial Kendall's $\tau$ correlation test~\citep{1996MNRAS.278..919A} shows that they are correlated when the influence of redshift is excluded, with $\tau = 0.30$ and $\sigma = 0.04$.

Different from the NLS1s detected by TGSS ~\citep{2019ApJ...879..107F}, the jet power of LoTSS detected NLS1s in this work is much lower than their disk luminosity (left panel of Figure~\ref{jet-disk}). The mean disk luminosity and jet power of LoTSS detected NLS1s are 10$^{44.84}$ erg s$^{-1}$ and 10$^{42.81}$ erg s$^{-1}$, respectively. The 77 TGSS detected NLS1s have similar mean value of disk luminosity (10$^{44.90}$ erg s$^{-1}$), but much higher mean value of jet power (10$^{44.22}$ erg s$^{-1}$). The mean value of jet production efficiency $P_{j}/L_{bol}$ for LoTSS detected NLS1s is 0.01. The right panel of Figure~\ref{jet-disk} shows the scatter between Eddington ratio and jet production efficiency. Weak negative correlation is found between them, with the correlation coefficient $\rho = -0.20$ and the chance probability $P = 0.004$.

LoTSS detected NLS1s show similar jet-disk connection with other jetted AGNs~\cite{2019ApJ...879..107F}. However, their lower jet production efficiency compared with the higher powered NLS1s suggests a potential evolution of jet activity between low powered jetted AGNs and their high powered counterparts, just like the low powered and classical type II radio galaxies of~\citeauthor{1974MNRAS.167P..31F} (FR IIs)~\citep{2019ApJ...879..107F}. This effect makes the unification model more complicated, and need more detailed comparison with large and completed samples of low powered jetted AGNs.

Flat spectrum radio quasars (FRSQs) and FR IIs are intrinsic the same type of AGNs in the unification model, where the observational differences are caused by the orientation effect~\citep{1995PASP..107..803U}. FSRQs are also suggested as unified sources with jetted NLS1s. They are hosted in the higher black hole mass systems than the later~\cite{2017FrASS...4....6F,2019ApJ...872..169P}.~\citeauthor{2019ApJ...879..107F} (\citeyear{2019ApJ...879..107F}) analysed the jet power and jet - disk connection for a sample of FSRQs, and found their mean values of Eddingtion ratio and jet production efficiency are $0.23$ and $0.04$, respectively. The corresponding values for the LoTSS detected NLS1s in this work are $0.31$ and $0.01$, respectively. These similar features on the accretion and jet activities indicate that the LoTSS detected NLS1s could be the low mass version of FR II-like, double lobe radio sources.

\section{Summary}
We explore the detection rate and radio-loud fraction with the LoTSS DR1 for the optically selected NLS1s from SDSS DR12. Our results show a relatively high detection rate of about 28\%, while a low radio-loud fraction of 1\%. The radio detected NLS1s show lower redshift than non-detected ones. The black hole mass, disk luminosity, and Eddington ratio are similar for both radio detected and non-detected sources. Radio loudness of radio detected NLS1s shows no dependence on central black hole mass, while weak correlations are found between radio loudness and disk luminosity, as well as Eddington ratio.

The radio luminosity, or the jet power of LoTSS detected NLS1s are more than one order of magnitude lower than that of the previous radio detected NLS1s samples. These LoTSS detected NLS1s show connections between radio emission and emission lines, which is common existing in other jetted AGNs. As the sensitivity and resolution of radio surveys get better, such as SKA and its pathfinders, the properties of the low luminosity radio AGNs, their places in the unification model of AGNs, and the origin of their radio emission need more attentions in the future.

\vspace{6pt}



\funding{This research was funded by National Natural Science Foundation of China (NSFC; grant number 11947099).}

\acknowledgments{We are grateful to the anonymous referees for their suggestive comments, which improve our manuscript greatly. This research made use of Astropy\footnote{http://www.astropy.org} (a community-developed core Python package for Astronomy \citep{2013A&A...558A..33A, 2018AJ....156..123A}) and Astroquery\footnote{http://astroquery.readthedocs.io/} \citep{2019AJ....157...98G}.}

\conflictsofinterest{The authors declare no conflict of interest.}


\reftitle{References}


\externalbibliography{yes}
\bibliography{bib}




\end{document}